# Machine Learning Reveals Composition Dependent Thermal Stability in Halide Perovskites

*Abigail R. Hering[1#], Mansha Dubey[1#], Elahe Hosseini[2#], Meghna Srivastava[1], Yu An[3] Juan-Pablo Correa-Baena[3,4], Houman Homayoun[2], and Marina S. Leite[1*]*

[1]Department of Materials Science and Engineering, UC Davis: Davis, 95616, USA
[2]Department of Electrical and Computer Engineering, UC Davis: Davis, 95616, USA
[3]School of Materials Science and Engineering, Georgia Institute of Technology: Atlanta, 30332 USA
[4]Department of Chemistry and Biochemistry, Georgia Institute of Technology: Atlanta, 30332 USA

*Corresponding author: mleite@ucdavis.edu
[#]These authors equally contributed to this work.

**Funding**: MSL and HH thank the financial support from DARPA (HR00112390144), and MSL also acknowledges the financial support from NSF (ECCS award #2023974) and the UC Davis Chancellor's Fellowship program. MD and MS thank the 2024 UC Davis Summer Graduate Student Researcher Award and the NSF Graduate Research Fellowship program (award 2036201), respectively.

**Abstract:** Halide perovskites exhibit unpredictable properties in response to environmental stressors, due to several composition-dependent degradation mechanisms. In this work, we apply data visualization and machine learning (ML) techniques to reveal unexpected correlations between composition, temperature, and material properties while using high throughput, *in situ* environmental photoluminescence (PL) experiments. Correlation heatmaps show the strong influence of Cs content on film degradation, and dimensionality reduction visualization methods uncover clear composition-based data clusters. An extreme gradient boosting algorithm (XGBoost) effectively forecasts PL features for ten perovskite films with both composition-agnostic (>85% accuracy) and composition-dependent (>75% accuracy) model approaches, while elucidating the relative feature importance of composition (up to 99%). This model validates a previously unseen anti-correlation between Cs content and material thermal stability. Our ML-based framework can be expanded to any perovskite family, significantly reducing the analysis time currently employed to identify stable options for photovoltaics.

**Main:**

Halide perovskites (HPs) have been the subject of intense research over the last decade due to their potential for improving semiconductor devices such as photovoltaics, light-emitting diodes, and radiation sensors.[1,2,3] These materials have several advantageous and tunable qualities, including state-of-the-art photovoltaics power conversion efficiencies exceeding 26%[4]. However, the Achilles heel remains their largely unpredictable behavior over time.[5] Their optoelectronic properties are especially dynamic in the presence of environmental stressors, such as temperature[6], humidity[7,8] oxygen[9], light[10], and bias[11]. Several degradation mechanisms have been proposed, including halide migration[12], phase transitions[13], and species degassing[14], but they are often convoluted and composition dependent.[15] Because of their fluctuating properties and variations between samples, a systematic, data-driven, machine-learning based approach is necessary[16,17] to accurately reveal correlations between compositions, stressors, and material properties.[18,19]

In the realm of materials for clean energy, artificial intelligence (AI) has recently allowed advancements in the fields of batteries, water splitting and fuel cells, and solar energy. Notable examples include forecasting remaining useful life in Li ion batteries,[20] determining whether solid electrolytes experience preferential stress failure through clustering algorithms,[21] screening of multidimensional solid-solution alloy space for water-splitting photocatalysts,[22] determining diffusion mechanisms of oxygen conductors for fuel-cells through machine learning (ML) trained interatomic potential molecular dynamics simulations,[23] and optimizing solar cells through dimensionality reduction of spectral features.[24] Concerning HPs for light-absorbing and -emitting devices, there has been growing interest in potential of machine learning (ML) to accelerate the understanding of this burgeoning class of material.[25] For instance, HPs photoluminescence response has been predicted with >90% accuracy for 50+ hours upon using a statistical modeling algorithm.[26] A multistep material screening based on density functional theory (DFT) has successfully identified 15 pseudo-halide molecule options (out of >$10^6$ molecules) that effectively passivate perovskite solar cells.[27] Self-supervised learning methods have also been implemented to map halide ion migration within HPs through non-destructive hyperspectral imaging[28]. Simultaneously, automated fabrication, characterization, and data analysis have become revolutionary in uncovering complex relationships in materials science through robotic synthesis[29,30] creation of large data libraries[31], and self-driving labs determining the most informative experiments.[32] Despite the progress in the field, it is still vastly unknown how the ultra-rich chemical composition space of the HPs is affected by temperature. Because of the hyperparameter space involved, reliable, temperature-stable HP optoelectronics are unlikely to be achieved using Edisonian methods.

Here, we elucidate and quantify the correlations between chemical composition within the $Cs_xFA_{1-x}PbI_yBr_{1-y}$ perovskite family and the effect of temperature on the material's optical behavior by combining automated, high-throughput *in situ* photoluminescence (PL) experiments and ML-driven analyses. We track the real-time PL response of the 10 perovskite films from this composition space upon submission to simultaneous temperature variations ranging from 15°C to 55°C. The 137,000 acquired spectra are used as input to train and test a series of ML algorithms. We generate correlation heatmaps that reveal a previously hidden connection between Cs content and material stability upon temperature variations. With dimensionality reduction through visualization techniques, we identify unique patterns between the input variables (composition, temperature, and time) and the following PL features: peak maximum ($PL_{max}$), peak area ($PL_{area}$), peak full-width-half-maximum ($PL_{fwhm}$), and peak location ($PL_{loc}$). We find that $PL_{max}$ and $PL_{area}$ are predicted with up to 98% accuracy through composition-agnostic ML models and with >75% accuracy in a generalized model that includes cation and halide content as inputs. Cumulative feature importance, which we extract from ensemble learning models, deconvolutes the weights of temperature, time, and composition on the optical response of each sample. Overall, the Cs content strongly dictates which samples will irreversibly degrade and which ones will recover from the cyclical temperature stress, where lower Cs content leads to higher thermal stability.

Our implemented workflow is displayed in Fig. 1. First, we select 10 compositions in the $Cs_xFA_{1-x}PbI_yBr_{1-y}$ family (Fig. 1Ai). We then implement a temperature profile that roughly mimics the extreme day and night temperatures of Sacramento, California, in the month of July, which are adapted from the National Weather Service (Fig. 1Aii). The HP samples are placed in a custom high-throughput PL sample chamber on a translational stage (Fig. 1Bi), where they experience identical environmental conditions. In this setup, we acquire 13,700 spectra in six days. The four PL features of interest are maximum intensity ($PL_{max}$), absolute area ($PL_{area}$), peak full-width-half-max ($PL_{fwhm}$), and peak location ($PL_{loc}$) (Fig 1Bii). During data preprocessing, these features are normalized and outliers are removed (Fig. 1Ci). The temperature profile and cation and halide content are integrated with the spectral data (Fig. 1Cii), and three data analysis and visualization techniques are employed (Fig. 1Ciii). Finally, seven ML algorithms are trained (Fig. 1Di) and tested (Fig 1Dii), first on each sample independently, and then as a generalized model which can predict the behavior of all 10 chemical compositions. The weighted feature importance of each PL attribute (Fig 1Diii) reveals which compositions are inherently unstable, and which show robustness against quick temperature variations.

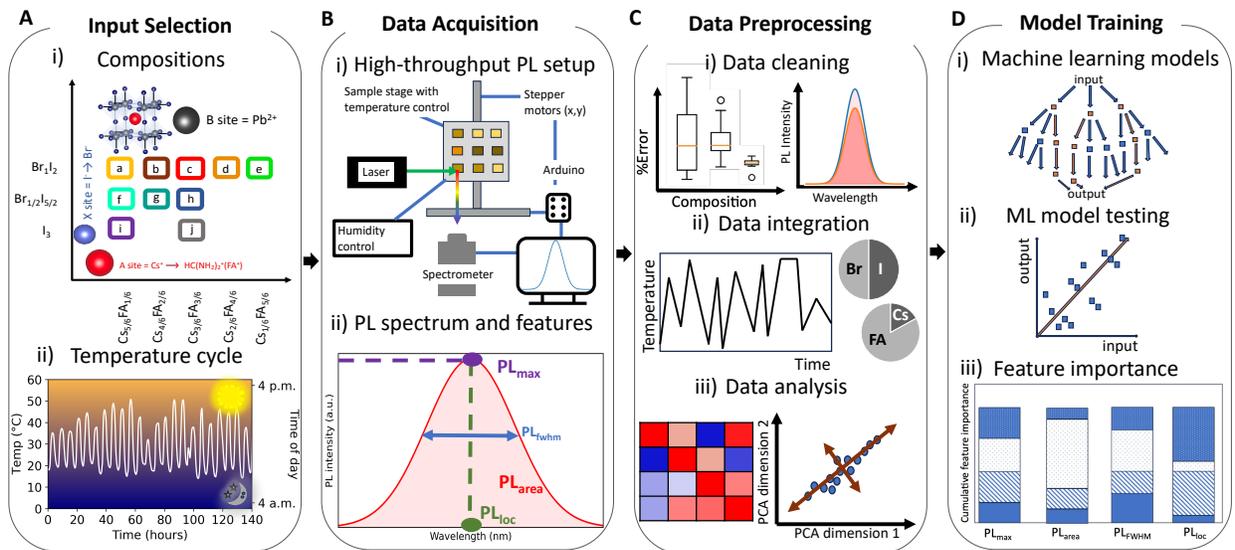

**Fig. 1**: **Workflow to reveal composition dependent properties in halide perovskites.** (**A**) (i) Selected $ABX_3$ perovskite compositions (labeled a-j), where A is a mixture of organic (formamidinium) and inorganic (cesium) cations, B is a lead cation, and X is a mixture of halide anions (bromine and iodine). (ii) Temperature profile during high-throughput experiments is 24 cycles between 15 °C and 55 °C, across 137 hours, mimicking the extreme summertime day and night cycles in Sacramento, California. Each day-night period is compressed into a 6-hour cycle. (**B**) (i) Data acquisition through (i) high throughput, *in situ* PL. (ii) Four key outputs: maximum PL intensity ($PL_{max}$), integrated PL area ($PL_{area}$), full-width half-maximum ($PL_{fwhm}$), and peak wavelength location ($PL_{loc}$). (**C**) Data preprocessing includes (i) data cleaning *via* normalization and removal of outliers, (ii) data integration, and (iii) data analysis using visualization techniques. (**D**) (i,ii) Model training and testing using tree-based machine learning algorithms. (iii) Weighted feature importance of each of the experimental inputs (time, temperature, and composition) on each PL feature output.

**Automated, high-throughput experiments**

High-throughput data acquisition is ideal for an AI-driven analysis, providing sufficient data to train ML algorithms with statistically significant and unbiased results. *In situ* photoluminescence (PL) spectroscopy captures the behavior of photogenerated charge carriers under various environmental conditions.[33] The primary features of a PL spectrum, $PL_{max}$, $PL_{area}$, $PL_{fwhm}$, and $PL_{loc}$, could be extrapolated to set guidelines for optimizing device parameters[34]. For example, the absolute PL intensity of a thin film is proportional to the maximum achievable solar cell open-circuit voltage ($V_{oc}$). Other PL metrics, such as $PL_{loc}$ and $PL_{fwhm}$, are indicative of the material's band gap and presence of any photoactive phases or band-edge states.[33] A PL spectrum contains key information that describes material properties and physical processes occurring in response to the environmental conditions. The significant reduction in time spent fabricating thin films, as compared to full devices, allows for more frequent testing on large sample sets.

We measure the PL of each HP thin film every six minutes, acquiring 13,700 spectra per sample (Fig. 2). The samples are subjected to temperature variations (Fig. 2Aii, white solid line for the profile), ranging from 15 °C and 55 °C. From these results, we see that the PL spectra for each sample modulate throughout the experiment in response to the temperature profile, however, some compositions exhibit more significant changes than others (see Supplementary Fig. 1 for initial spectra). For example, sample *a* initially has two peaks present (Fig. 2Ai). The peak at the shorter wavelength, 669 nm, corresponds to the true perovskite peak for this composition, whereas the one at 710 nm refers to a poorly mixed $Cs_{1-x}PbBr_{1-y}I_y$ phase, due to the known relative immiscibility of CsBr in dimethyl sulfoxide (DMSO) and the thermodynamic instability of the pure perovskite phase.[35] The longer-wavelength peak diminishes in intensity and eventually disappears over the course of the experiment (Fig. 2Aii). The eventual absence of this peak can be attributed to entropic mixing of the various phases at elevated temperatures in an inert atmosphere. Previous studies have shown that phase segregation in HPs can be reversible upon the removal of external stimuli.[36,37]

Variations in the shape and evolution of PL spectra demonstrate the discrepancies across HP compositions (see Fig 2Ai-Ji). Shoulders in the PL spectra are observed for samples *b*, *c*, *d*, *e*, and *f* at 650 nm, which are likely the result of a segregated Br-rich phase[12]. Conversely, I-rich compositions (*g, h, i* and *j*) show no secondary peaks or shoulders, indicating a well-mixed perovskite phase. The $PL_{fwhm}$ increases and decreases with each temperature cycle, due to the changes in the concentrations of halide vacancies and interstitial defects at various temperatures.[38] These defect states have been shown to form near the band edges as shallow trap states, and they are prominent enough to modify the spectra intermittently.[39] There is no observable change for sample *g* at the conclusion of the experiment, indicating excellent stability against thermal fluctuations. This sample recovers its initial $PL_{max}$ maintains a relatively constant $PL_{fwhm}$ throughout the thermal cycling. Samples *a*, *b*, *c*, *f* and *i*, with Cs content ≥50%, have the largest reduction in $PL_{max}$ at the end of the experiment. Typically, low amounts of Cs (17%-33%) added to FA result in more stable perovskites[40] by altering the structure of the lattice in a favorable way.[41] An ideal tolerance factor of 1.0 is achieved with small amounts of Cs, but excessive Cs induces strain on the lattice, ultimately leading to segregation of the photoinactive $\delta$-phases.[42] Overall, the real-time tracking of PL features provides key insights into the chemical processes occurring within these materials.

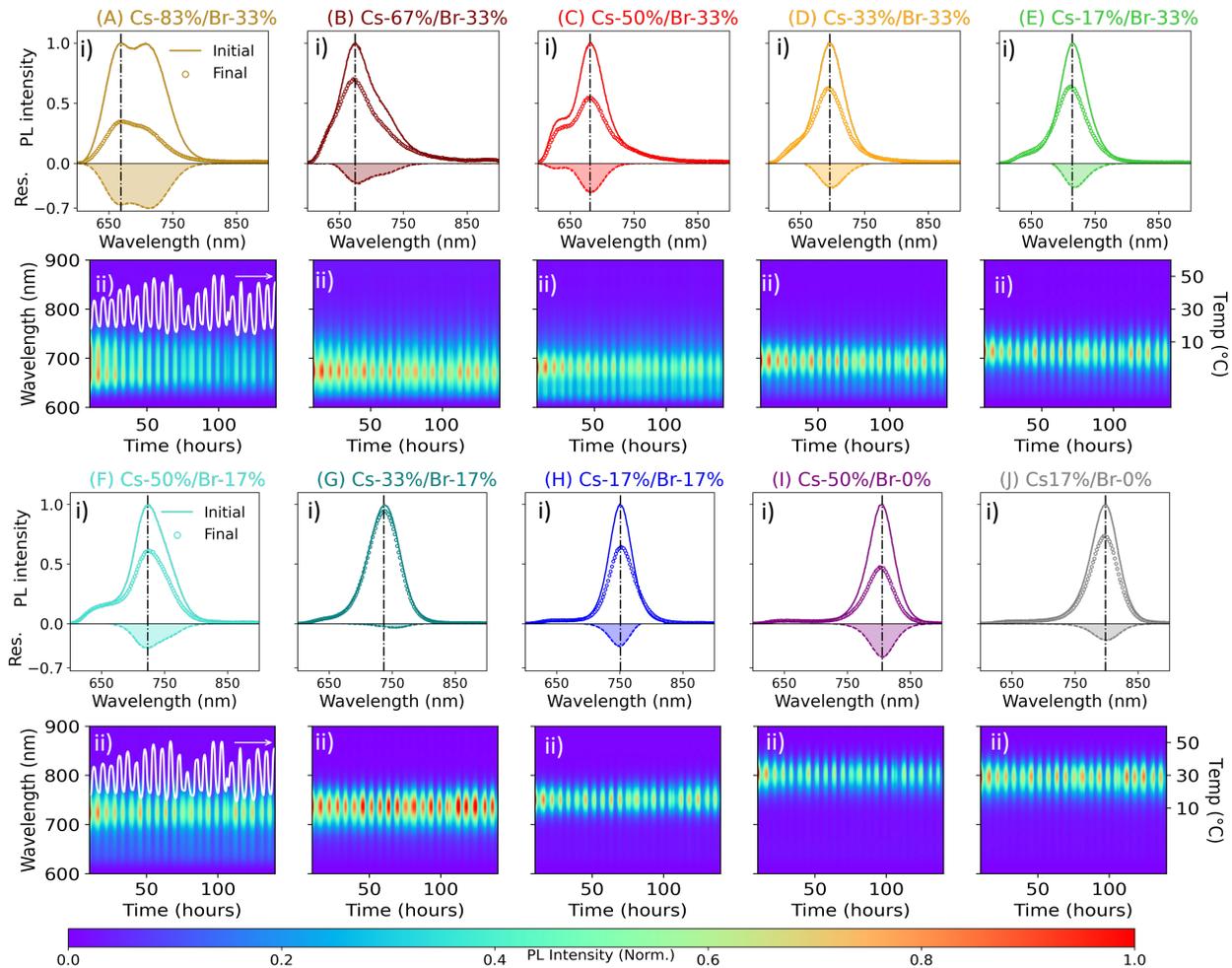

**Fig. 2**: **Tracking real-time materials change through *in situ* PL**. (**A**i-**J**i) Initial (solid line) and final (open circles) PL spectra of each perovskite composition, referring to t=0 hours and t=137 hours, respectively. Both are normalized to the maximum value of the initial spectrum and are acquired at room temperature. The residual shows relative changes in shape and intensity between the PL spectra. The dashed line marks the center wavelength of the initial PL spectrum. (**A**ii-**J**ii) Heatmaps of 13,700 PL spectra for all samples where 1 (red) and 0 (blue) correspond to the maximum intensity of the first spectrum for each sample and to the baseline, respectively. The temperature profile is shown in white in subplots (**A**ii) and (**F**ii).

**Statistical analysis and data visualization**

In multivariate data analysis, correlation heatmaps serve as a powerful tool for displaying the strength and direction of relationships among variables. Pearson correlation coefficient (PCC), a statistical metric that quantifies the linear relationship between two variables, is used as a standard measure.[43] The correlation heatmaps (Fig. 3A and B) reveal the relationship between each of the inputs, temperature and time, and the PL features of samples *a* and *g* (see Supplementary Fig. 8 for all heatmaps). Fig. 3C shows the PCC between the temperature, time, cation, and halide content (represented by Cs and Br, respectively), and the output PL features, across the entire dataset. Overall, temperature is anti-correlated with $PL_{max}$, $PL_{area}$, and $PL_{loc}$, and correlated with $PL_{fwhm}$.

Sample *a* shows that temperature has weaker correlations (<0.69) with PL features than those of sample *g* (>0.90), which has very cyclical, reversible behavior. Time has much stronger correlations with the PL features of sample *a* (>0.36) than with sample *g* (<0.14), indicating that sample *a* changes and degrades over time. Time has weak correlations with sample *g*'s PL features, indicating that this HP composition fully recovers from its environmental stressing after each cycle. When considering the collective response of all compositions (Fig. 3C), temperature has a moderate correlation with PL features in the averaged dataset (<0.67), primarily attributed to changes in peak shape occurring across the different samples and, thus, different degrees of material degradation. Overall, the negative correlations with $PL_{max}$ (-0.44) and $PL_{area}$ (-0.44), in addition to the positive correlation with normalized $PL_{fwhm}$ (0.67), clearly demonstrate the adverse effects of thermal cycling on most HP compositions. There is minimal temperature effect on $PL_{loc}$ (-0.03), as no phase transitions occur in this temperature regime. The Br content has a strong correlation (-0.96) with $PL_{loc}$, which is representative of the material's band gap[44], and Cs concentration shows a modest correlation with $PL_{loc}$ (-0.54). This result is a direct consequence of the fact that the valence band maximum and conduction band minimum are mainly derived from the overlap of Pb and Br/I orbitals, which deviate based on size of the A site cation. Br shows a stronger correlation with $PL_{max}$ than Cs (-0.50 vs. -0.44), indicating that the presence of mixed halide phases can introduce more defects in the system, leading to additional non-radiative recombination pathways (and $PL_{max}$ reduction). This outcome also indicates that samples with modest amounts of Cs and Br have higher radiative recombination. The correlations of composition with $PL_{area}$ are slightly weaker than with $PL_{max}$, due to irregular peak shapes.

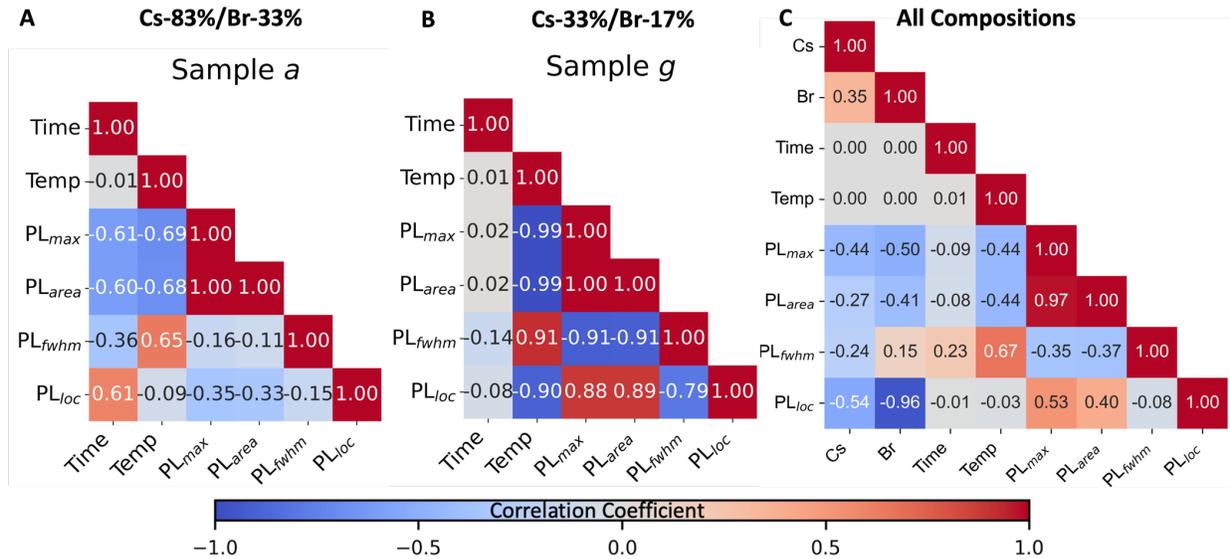

**Figure 3**: **Pearson correlation coefficients between composition, time, temperature, and PL features**. Heatmap of parameters for (**A**) Cs-83%/Br-33% (sample *a*), (**B**) Cs-33%/Br-17% (sample *g*), and (**C**) all 10 perovskite compositions. A value of 1 and -1 indicate a perfect positive and negative correlation, respectively. A value of 0 refers to no correlation between the variables.

Principal Component Analysis (PCA) and t-Distributed Stochastic Neighbor Embedding (t-SNE) are employed to visualize the characteristics of the dataset. These dimensionality reduction techniques integrate input features of compositional elements (Cs, Br) and environmental factors (temperature and time) and enable the high-dimensional PL attributes to be represented in a two-dimensional space. In turn, this analysis facilitates the identification of patterns and relationships between the input variables and the experimental results. Figure 4A illustrates the distribution of the samples in the space defined by the PCs, with each point representing a single sample's location within this novel coordinate system. Here, the colors represent each sample, offering a visual insight into how these factors correlate with the PCs. Figure 4B illustrates how samples are distributed in this two-dimensional t-SNE space, with the spatial proximity of points suggesting similarities in the PL properties under corresponding input parameters. This analytical visualization can reveal complex dependencies and behaviors within the PL property space, informed by both material composition and environmental factors. Further, each composition cluster can be distinguished as they refer to a unique response behavior.

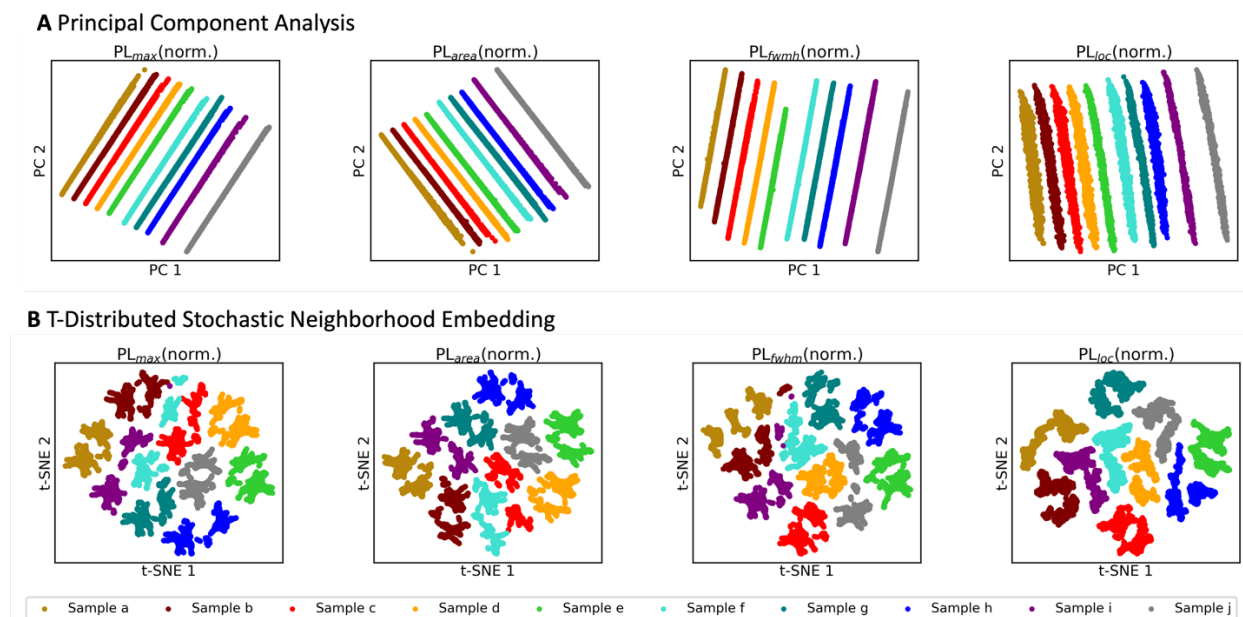

**Figure 4: Data visualization *via* dimensionality reduction**. (**A**) Principal Component Analysis (PCA) showing linear, global data distribution for the normalized PL features. (**B**) t-Distributed Stochastic Neighborhood Embedding (t-SNE) with nonlinear, local data distribution. In each scatterplot, the clusters corresponding to 10 perovskite compositions within the dataset for unique and distinguishable patterns.

**Machine-learning-based analysis of perovskites thermal stability**

We employ a comprehensive ML paradigm to predict PL characteristics, with two approaches: ten composition agnostic models and a generalized, composition-dependent model. In each case, the data for all HPs is normalized, allowing more robust convergence during model training. For the first approach, we train seven different ML algorithms (Bagging Regressor, Decision Tree Regressor, Extra Trees Regressor, Gradient Boosting, K-Nearest Neighbors Regressor, Random Forest Regressor, and XGBoost), individually for each of the 10 compositions, using an 80:20 split for training and testing data. The inputs to these models are time, and temperature, and the outputs are the four PL features. The Root Mean Square Error (RMSE) is used to directly compare results for each sample across all PL properties, showing that performances of the ML models vary significantly between the PL characteristics. XGBoost consistently demonstrates prediction capabilities with average accuracy >85% (see Supplementary Fig. 10 and Supplementary Tables 1-4), which highlights the effectiveness of the gradient boosting technique in dealing with complex nonlinear relationships. Tree-based algorithms are less computationally expensive than neural networks, and have a lower risk of overfitting, due to tree pruning, which makes them ideal for analyzing our multivariable data space.

The ability to accurately predict the behavior of HPs in a variety of compositions and environments is essential for the rapid development and application of energy-saving devices using HPs.[45] Thus, we realize a robust, generalized model to predict the PL behavior of new compositions accurately, reducing the need for extensive experimental testing. A generalized model, validated by Leave-One-Group-Out (LOGO) cross-validation, is a rigorous method where one sample is kept as a test set while the remaining ones are used for training, providing scalability, efficiency, and increased predictive power. This process is repeated until each unique sample is used as a testing set, ensuring that the model's performance is evaluated across all independent sample groups (see Supplementary Fig. 11 and Supplementary Tables 5-8 for RMSE results), with a total accuracy average >75%.

For deeper insight into the relative impact of various factors on PL properties, we conduct a feature importance analysis[46,47] of the XGBoost algorithm. Fig. 5 displays the cumulative feature importance of time and temperature on PL features of samples *a*, *g*, and the generalized model (see Supplementary Fig. 12 for all results). Sample *a* shows stronger dependence on time than temperature (>0.5) for all PL features. We attribute this disparity to the cyclic fatigue and inability to recover apparent in sample *a* (see Fig. 2Ai). Temperature dominates all PL characteristics of sample *g* (Fig. 5B), while time has low importance (<0.2), which indicates that the Cs-33%/Br-17% composition responds cyclically to the temperature stressor and recovers its properties after each cycle. Fig. 5C displays feature importance prediction across all compositions for the generalized model approach. This analysis considers the relevance of time, temperature, Cs, and Br for predicting PL features. Notably, the cation and halide content are the most significant factors, contributing substantially to the predictions of PL outputs. Temperature still affects $PL_{max}$, $PL_{area}$, and $PL_{fwhm}$, with the latter showing the strongest effect. As expected, $PL_{loc}$ is insensitive to the temperature variations tested here, and depends most strongly on Br content. Overall, these proof-of-concept results show that Cesium content plays a significant role in dictating the robustness of the materials' optical properties in response to thermal modulations.

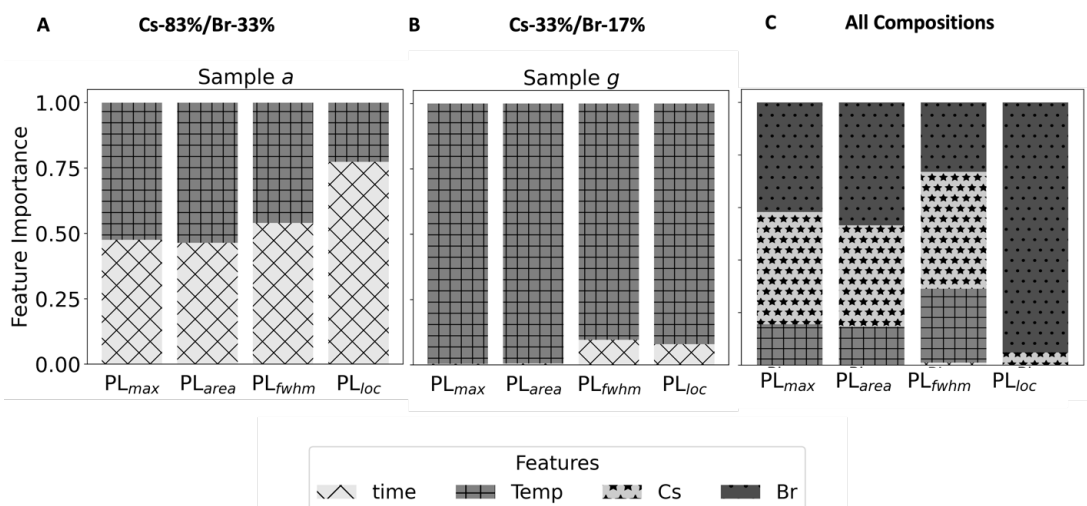

**Figure 5**: **Cumulative feature importance in machine learning model XGBoost predictions.** Feature importance as a function of PL metrics for (a) Cs-83%/Br-33% (sample *a*), (b) Cs-33%/Br-17% (sample *g*), and all compositions (c). For (c), composition dominates all PL outputs, while time has little importance in the generalized model.

**Discussion and outlook**

We unraveled the correlations between cation and halide compositions, temperature, and time on four key PL features (peak maximum intensity, peak location, full-width half maximum, and integrated peak area) using ML. Our automated and *in situ* PL measurements enabled the acquisition of 137,000 spectra, sufficient input for an ML-driven analysis, while stressing a series of Cs-FA halide perovskites. We found trends that clearly demonstrate the detrimental effects of temperature cycling on the majority of HP films, as seen by a decrease in PL intensity. However, sample degradation is largely composition dependent, and the most stable ones present modest Cs and Br content. The correlations and feature importance of each sample's composition, temperature, and PL properties elucidate the role of cation content on perovskite's ability to resist and recover from exposure to a relevant environmental stressor: temperature. Key PL features were predicted with >85% accuracy in composition-agnostic models and >75% accuracy in composition-dependent ones. The ML-based analysis enabled us to uncover a strong anti-correlation between Cs content and material thermal stability. Both ML model approaches can be successfully implemented to any HP compositions with a range of input variables that extend beyond temperature. Our ML-based framework could be used to quantify the combined effects of stressors on material degradation and, ultimately, on device performance.

**Methods**

Sample Fabrication: The perovskite sample fabrication process follows the recipe from references [35], [26]. The precursor chemicals, formamidinium iodide, formamidinium bromide, lead iodide, lead bromide, cesium iodide, and cesium bromide are purchased from Dyenamo (FAI and FABr), TCI America ($PbI_2$ and $PbBr_2$), and Sigma-Aldrich (CsI and CsBr). The solvent and antisolvent, dimethyl sulfoxide (DMSO) and chlorobenzene (CB), respectively, are from Arcos Organics. The substrates, fluorine-tin-oxide (FTO) coated glass (10 ohms/sq) are purchased from Kintec. The precursor solutions of $CsPbBr_3$, $CsPbI_3$, $FAPbBr_3$, and $FAPbI_3$ are prepared as 0.4M solutions in DMSO in a nitrogen glovebox. They are then mixed in the appropriate ratios to achieve the various $Cs_xFA_{(1-x)}Pb(Br_yI_{(1-y)})_3$ stoichiometries. The perovskite solutions are then spin coated onto clean FTO substrates at 500 rpm with an acceleration of 250 rmm/s for 10 seconds. The second spin step is set at 2,000 rpm with an acceleration of 1,000 rpm/s for 70 seconds, with CB antisolvent dropped in the last five seconds. Finally, the films are annealed on a hot plate at 65 °C for 10 minutes. All samples are stored in nitrogen environment and transferred to the *in situ* PL setup without exposure to ambient conditions.

Experimental Setup: The samples are placed in a home-built, high-throughput PL experimental chamber (Figure 1Bi) and submitted to identical environmental conditions. An Arduino controls two stepper motors, so that one sample is illuminated by the stationary excitation laser (a Vortran 532 nm dioide laser at constant 1 mW power) at a time, and the experiment can be run fully autonomously. The temperature, humidity, and full PL spectra are collected for each sample every six minutes throughout a six-day experiment (Fig. 1Aii). The detector is a Princeton Instruments HRS-300 Spectrometer, which saves each spectrum as an output file that has a corresponding time stamp, temperature, and humidity value. Supplementary Fig. 1 displays the original PL spectra for each sample.

Temperature is controlled *via* a Linkam PE120 heating element, modified to attach to the back of the sample chamber. An Oasis cooling system is used to ramp down temperatures. An Adafruit SH40 temperature and humidity sensor is attached inside the sample chamber, for *in situ* environmental logging. The temperature accuracy is ±0.2 °C and the relative humidity accuracy is ±1.8%. The PL features, peak maximum intensity ($PL_{max}$), integrated peak area ($PL_{area}$), peak full-width half maximum ($PL_{fwhm}$), and peak center location ($PL_{loc}$) are normalized for the ML analysis for better comparison between samples. The full variation of each output throughout the experiment is displayed in Supplementary Figs. 2-5 for all compositions. The most prominent changes to each HP occur within the first three cycles of the experiment (Supplementary Fig. 6), where the difference between samples undergoing changes, such as halide remixing (sample *a*) and stable, robust samples (sample *g*) are evident.

Data Cleaning: Careful data processing is essential to extract and quantify correlations between the different compositions and the PL features. To process the data, we first remove outliers outside of four times the interquartile range (Supplementary Fig. 7). The identification of outliers is a critical aspect of data preparation for ML for ensuring the robustness of the models, as these data points can significantly impact the accuracy of training step and ultimate predictions. Second, we normalize the data, so that relative changes across each sample are treated equally by the models and for easy visualization of the distribution of each of the four features: $PL_{max}$, $PL_{area}$, $PL_{fwhm}$, and $PL_{loc}$.

The box plots provide a visual summary of these distributions, where the central line in each box represents the median value, the edges of the box are the 25$^{th}$ and 75$^{th}$ percentiles (the interquartile range, IQR). The whiskers extend to 4x the IQR (Supplementary Fig. 7), which is a conservative threshold that ensures only the most extreme variations in PL are classified as outliers (denoted by diamond symbols). Outliers in PL property distributions may indicate experimental errors, sample inconsistencies, or data processing anomalies, thus, they are removed from the ML analysis.

Data Visualization: PCA orthogonally transforms the original variables into a new set of linearly uncorrelated variables termed principal components (PCs). The first PC captures the maximum variance present in the data, and each subsequent component has the highest variance possible under the constraint of being orthogonal to the preceding ones. The methodology involves standardizing the dataset, calculating the covariance matrix, and then extracting the eigenvalues and eigenvectors of this matrix, which, in turn, dictate the magnitude and direction of the new space, respectively. By projecting the original data along these new axes, PCA provides a means to reduce the dimensionality of the dataset. Supplementary Fig. 9A illustrates the distribution of the samples in the space defined by the PCs, with each point representing a single sample's location within this novel coordinate system. Here, the colors indicate the value of each PL property, offering a visual insight into how these factors correlate with the PCs. A second visualization technique, t-SNE, is an advanced technique designed for representing high-dimensional datasets by reducing them to low-dimensional spaces for ease of exploration and interpretation[48]. It has been successfully adopted in situations ranging from medical monitoring [49] to malware detection [50], where computer-engineering-based models have enabled data analysis intractable for humans. Briefly, this method proceeds by measuring the similarities between pairs of instances in the high-dimensional space and translating these into joint probabilities. t-SNE then iteratively adjusts the coordinates in the lower-dimensional space in a way that similar and distinct objects are modeled by nearby and distant points, respectively, effectively preserving the local structure of the data. The optimization is guided by the minimization of the Kullback–Leibler divergence between the joint probability distributions in both the original and reduced spaces[51]. The clusters shown in Supplementary Fig. 9B informs mathematical relationships between the input and output features and, thus, motivate further ML analysis.

Machine Learning Algorithms: We employ and compare seven ensemble learning methods such as Bagging Regressor[52], Random Forest Regressor[53], Extra Trees Regressor[54], Gradient Boosting[55], and XGBoost[56] and single model methods such as Decision Trees Regressor[53] and K Nearest Neighbors (KNN) Regressor[57] in the ML analysis. Briefly, ensemble learning is an ML technique that aggregates the predictions of several base models by averaging or voting to produce a more robust and accurate final prediction. These techniques are much less computationally intensive than neural networks. Bagging Regressor averages the outputs of multiple models trained on different subsets of the training data to reduce variance and improve model stability and accuracy. Decision Trees Regressor uses a tree-like structure of decisions that recursively breaks down a dataset into subsets based on the values of its features. Extra Trees is a variant of Random Forest that constructs a multitude of decision trees with random splits in nodes instead of optimum split to capture more diverse patterns. Gradient Boosting combines the predictions of multiple weak learners sequentially. In each iteration, the algorithm aims to minimize the loss function of the previous model using gradient descent. KNN regression is a non-parametric supervised method that predicts the output value by averaging the 'K' nearest training examples in the feature space.

Random Forest constructs multiple decision trees and averages their predictions to provide durability against overfitting. Finally, XGBoost is a scalable and optimized distributed gradient boosting library. The results of these seven ML algorithms are displayed in Supplementary Fig. 10 and Supplementary Tables 1-4 for the composition-agnostic models. These models are trained and separately for each composition and each PL feature, for a total of 280 independent ML models. The relative feature importance for all samples *a-j* are shown in Supplementary Fig. 11. Note that samples *a* and *g* are also displayed in Figure 5, as they represent the extreme ends of sample stability. The results of the generalized model, with the Leave-One-Group-Out (LOGO) cross-validation are displayed in Supplementary Figure 12 and Supplementary Tables 5-8. These models are trained for each PL feature, for a result of 28 independent ML models.

**Data and materials availability:** All relevant data are available in the main text or the supplementary materials. All other data supporting findings are available to be shared upon request.

**Code availability:** All code is available to be shared upon request.

**Acknowledgements:** The authors acknowledge T. Deppe and Dr. J. N. Munday for fruitful discussions. We thank D. Hemer from UC Davis SHOPS (a joint effort by Crocker Nuclear Lab and the Biological and Agricultural Engineering Department) for technical assistance with our customized sample holder. MSL and HH thank the financial support from DARPA (HR00112390144), and MSL also acknowledges the financial support from NSF (ECCS award #2023974) and the UC Davis Chancellor's Fellowship program. MD and MS thank the 2024 UC Davis Summer Graduate Student Researcher Award and the NSF Graduate Research Fellowship program (award 2036201), respectively.

**Author Contributions**: ARH and MD analyzed and interpreted the photoluminescence data and drafted the manuscript. EH conducted the machine learning analysis and contributed to the manuscript draft. MS collected the designed the experiment and collected the data. YA fabricated the perovskite films. ARH, MD, and EH created the figures. MSL, HH, and JPCB supervised the project. MSL conceived the idea. All authors contributed to the discussions and revisions of the manuscript.

**Ethics Declaration:** Authors declare that they have no competing interests.